\documentclass{nic-series}

\begin{document} 

\title{The State of Factoring on Quantum Computers}

\author{Dennis Willsch \inst{1,2} \and 
        Philipp Hanussek \inst{1,2} \and
        Georg Hoever \inst{2} \and
        Madita Willsch \inst{1,3} \and \\
        Fengping Jin \inst{1} \and
        Hans De Raedt \inst{1} \and
        Kristel Michielsen \inst{1,3,4}}

\authortoc{D. Willsch, P. Hanussek, G. Hoever, M. Willsch, F. Jin, H. De Raedt, K. Michielsen}

\institute{J\"ulich Supercomputing Centre, Institute for Advanced Simulation, 
  Forschungszentrum J\"ulich, 
  52425 J\"ulich, Germany\\
    \email{\{d.willsch, m.willsch, f.jin, h.de.raedt, k.michielsen\}@fz-juelich.de}
  \and
  FH Aachen University of Applied Sciences, 52066 Aachen, Germany\\
    \email{philipp.hanussek@alumni.fh-aachen.de, hoever@fh-aachen.de}
  \and
  AIDAS, 52425 J\"ulich, Germany
  \and
  RWTH Aachen University, 52056 Aachen, Germany
}

\maketitle

\begin{abstracts}
We report on the current state of factoring integers on both digital and analog quantum computers.
For digital quantum computers, we study the effect of errors for which one can formally prove that Shor's factoring algorithm fails.
For analog quantum computers, we experimentally test three factorisation methods and provide evidence for a scaling performance that is absolutely and asymptotically better than random guessing but still exponential.
We conclude with an overview of future perspectives on factoring large integers on quantum computers.
\end{abstracts}

\section{Introduction}
\label{willsch_sec_introduction}

The integer factorisation problem (IFP) is one of the oldest and most fascinating problems in mathematics~\cite{Bressoud,Lehman1974FactoringLargeIntegers}. It is defined as the problem of finding a non-trivial divisor of a composite integer $N$. Besides its historical significance, the IFP is of central importance to everyday data and communication security, in the sense that the security of common encryption systems and protocols in use is based on the difficulty of solving the IFP for large integers. The latest record is the factorisation of the 829-bit number RSA-250 from the RSA factoring challenge~\cite{Boudot2020FactoringRSA250}, involving a 32M-hour allocation on the JUWELS supercomputer. The best-known algorithms~\cite{Boudot2020FactoringRSA250,lenstra1993developmentOfTheNumberFieldSieve,Kleinjung2010FactorizeRSA768Modulus} to solve the IFP on conventional computers scale (sub)exponentially in the number of bits of the integer $N$. For this reason, cryptosystems like RSA~\cite{RSA1978RSA}---currently using integers $N$ with 1024, 2048, or 4096 bits---are still secure.

Quantum computers (QCs) are an emerging technology that  promise a breakthrough in the solution of the IFP. We distinguish between \textbf{digital} and \textbf{analog} QCs. On an ideal digital QC, Shor's algorithm~\cite{shor1994factoring,ekert1996quantumalgorithms,shor1997algorithm} can solve the IFP with time and space complexity that is polynomial---not exponential---in the number of bits of $N$. 
However, so far only very small integers $N\le35$ have been successfully factored~\cite{Vandersypen2001ExperimentalShorNMR,MartinLopez2012ShorExperimentQubitRecycling,Monz2016ShorOnTrappedIons,Amico2019ShorOnIBMQ} with Shor's algorithm on a digital QC\footnote{Note that there exist many claims of factoring larger integers on digital QCs, but the underlying experiments often rely on a certain kind of oversimplification~\cite{Smolin2013OversimplifyingShorFactoring} that makes them equivalent to coin flipping.
Even for $N=15,21,35$, one can argue that the explicitly compiled quantum circuits might not have been found without previous knowledge about the answer to the IFP.}. 
By executing Shor's algorithm on a QC simulator using 2048 GPUs of JUWELS Booster, the largest integer that could be factored is the 39-bit number $N=549\,755\,813\,701=712\,321\times771\,781$~\cite{Willsch2023Shor} (see Table 3 in Ref.~\citen{scholten2024IBMAssessingBenefitsAndRisksOfQuantumComputing} for an overview). 

For analog QCs, several alternative approaches to solving the IFP exist
\cite{Peng2008QuantumAdiabaticAlgorithmForFactorization,Schaller2010FactoringAdiabaticQuantumAloorithm,Andriyash2016BoostingIntegerFactorizationDWave,Dridi2017PrimeFactorizationDWave,Jiang2018QuantumAnnealingForPrimeFactorization,Peng2019FactoringLargeIntegersDWave,Mengoni2020BreakingRSAWithDWave2000Q,Wang2020PrimeFactorizationParemeterOptimizationIsingAnnealer,Lanthaler2023ReversibleParityGatesForIntegerFactorization,ding2023FactoringOnDWaveLocallyStructuredEmbedding}.
Analog QCs hold the current record of the largest non-trivial integer factorisation by QC hardware, namely the 23-bit integer $N=8\,219\,999=251 \times32\,749$ factorised by a D-Wave quantum annealer~\cite{ding2023FactoringOnDWaveLocallyStructuredEmbedding}.
Although a polynomial scaling of factorisation by analog QCs has been suggested numerically~\cite{Peng2008QuantumAdiabaticAlgorithmForFactorization}, an exponential scaling is considered more likely.
In this article, we show experimental evidence for the latter.

This article is structured as follows. 
Section \ref{willsch_sec_digitalqc} focuses on solving the IFP with digital QCs. We review the main ideas of Shor's algorithm, its large-scale simulation on JUWELS, and future perspectives of factoring on digital QCs.
Section~\ref{willsch_sec_analogqc} discusses three methods of factoring on analog QCs. In this section, we also present results of implementing these methods on quantum annealers. 
Section \ref{willsch_sec_conclusions} contains our conclusions.

\section{Digital Quantum Computers}
\label{willsch_sec_digitalqc}

A digital QC---also known as a \emph{gate-based} or \emph{universal} QC\cite{NielsenChuang}---is a machine consisting of individually controllable \emph{quantum bits} (qubits). A qubit is defined as a superposition of the classical-bit states ``0'' and ``1'' and is commonly written as $\alpha\vert0\rangle+\beta\vert1\rangle$ with $\alpha,\beta\in\mathbb C$. Crucially, when a qubit is measured at the end of a computation, one always obtains one of the two classical-bit states, namely either ``0'' with probability $\vert\alpha\vert^2$ or ``1'' with probability $\vert\beta\vert^2$. Hence, every QC is a \emph{probabilistic} machine. In a digital QC, each individual qubit (and certain combinations of multiple qubits) are individually operable, and these operations are called \emph{quantum gates}. A digital QC is called \emph{universal}, because in principle, each program for conventional computers can be mapped to a combination of quantum gates with only polynomial overhead (note that this does not imply that everything will run faster on a QC---currently, only a few algorithms with a proven speedup are known). 

The current three most promising technologies for digital quantum processing units (QPUs) are superconducting circuits, neutral atoms, and trapped ions. With superconducting circuits, IBM has manufactured a 1121-qubit QPU~\cite{Castelvecchi2023IBM1121qubits}, and Google has demonstrated \emph{quantum error correction} below the surface code threshold on a 105-qubit QPU~\cite{google2024quantumerrorcorrection105qubits}. With neutral atoms, QuEra has built a logical QPU with 280 physical qubits~\cite{Bluvstein2024NeutralAtomLogicalQuantumProcessor}. Finally, trapped ion QPUs produced by Quantinuum have achieved the best gate performance and an all-to-all connectivity~\cite{daSilva2024QuantinuumH2,decross2024QuantinuumH2Upgrade}.
Pioneering European companies producing digital QPUs are IQM focusing on superconducting circuits~\cite{Ronkko2024IQMSpark} and eleQtron focusing on trapped ions~\cite{Piltz2014eleQtronQPU}, both of which are being installed for provision at JSC. However, it is important to realise that all existing digital QPUs are still noisy prototypes, meaning that they can usually not compete with conventional (super)computers for most application problems.

\subsection{Shor's Factoring Algorithm}
\label{willsch_sec_shor}

\begin{figure}[t]
\begin{center}
\includegraphics[width=\textwidth]{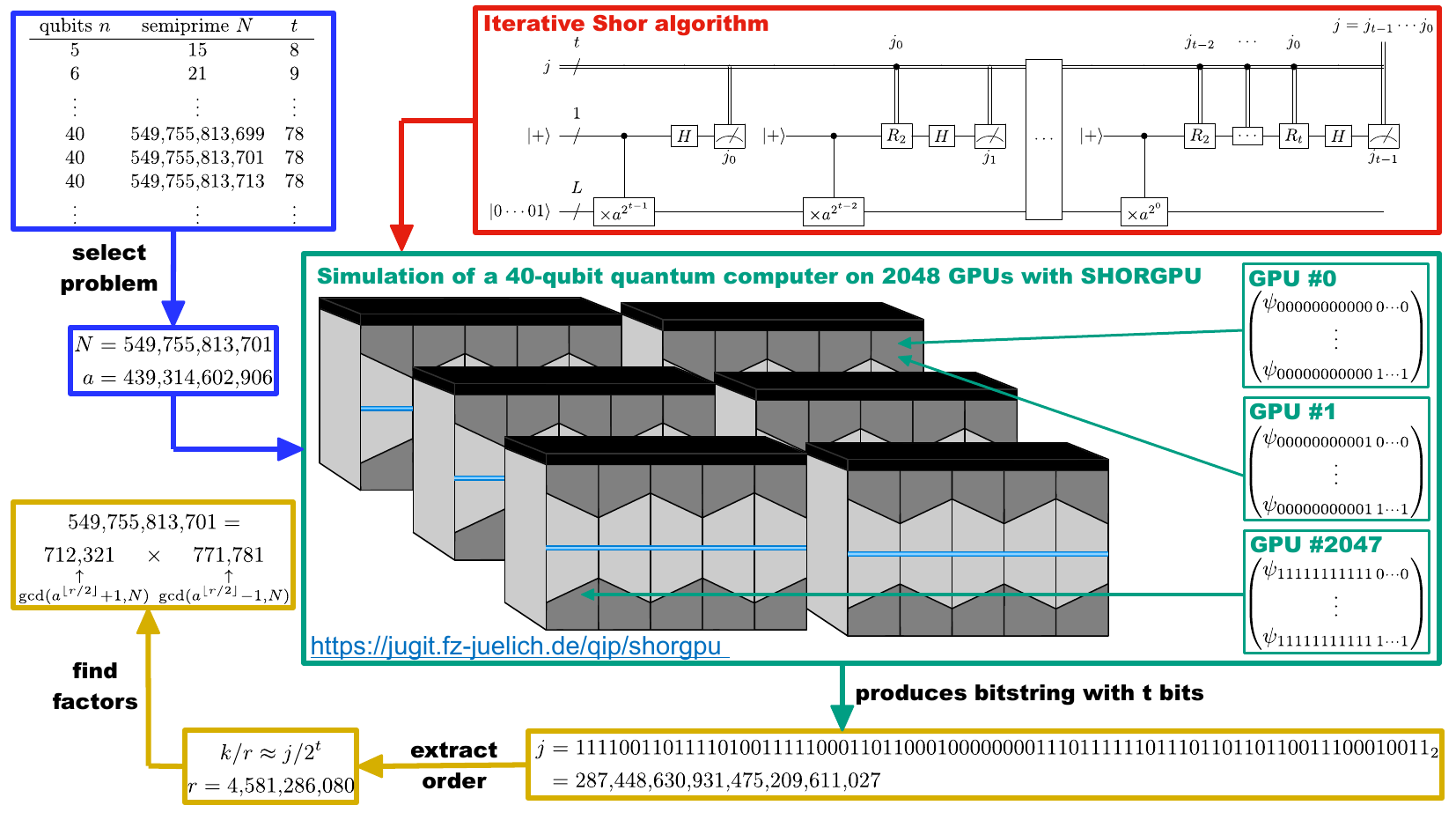}
\caption{\label{willsch_fig_scheme}
Schematic of performing Shor's factoring algorithm. First, one selects an $L$-bit semiprime $N$ to factor and a random $a$ (blue). Then, one executes the quantum gates of Shor's algorithm (red), using either a working digital QC or a large-scale simulation on multiple GPUs with \texttt{shorgpu}~\cite{shorgpu}, which yields a bitstring $j_0j_1\cdots j_{t-1}$ (green). Note that the iterative quantum circuit needs $L+1$ qubits to simulate an $L$-bit factoring scenario. Finally, the integer $j$ corresponding to the bitstring is post-processed, which yields with high probability a factor of the semiprime $N$ (yellow). Further details are given in Ref.~\citen{Willsch2023Shor}.
}
\end{center}
\end{figure}

Peter Shor proposed an algorithm to solve the IFP with an exponential speedup on an ideal  digital QC in 1994~\cite{shor1994factoring,shor1997algorithm}, a result which arguably sparked most of the community's interest to build a digital QC until the present moment. To explain Shor's factoring algorithm, we consider the factorisation of a semiprime $N=p\times q$, i.e., a composite integer $N$ with two unknown, non-trivial prime factors $p,q>2$. 
The algorithm consists of four steps that are schematically shown in Fig.~\ref{willsch_fig_scheme}:
\begin{enumerate}
    \item \textbf{Parameter Selection} (blue): Choose a random integer $a$ with $2\le a <N$ and greatest common divisor $\mathrm{gcd}(a,N) = 1$.\footnote{Note that if the greatest common divisor is not 1, it would have to be either $p$ or $q$, and the problem would have been solved by accident---which is very unlikely for large $N$. The greatest common divisor can be computed efficiently with the Euclidean algorithm.}
    \item \textbf{Quantum Algorithm} (red $+$ green): Execute the quantum gates of Shor's algorithm on a digital QC. The result of the QC are $t$ bits $j_0j_1\cdots j_{t-1}$, which make up the binary representation of an integer $j$. The number of bits $t$ is usually twice as large as the number of bits in $N$. Note that, in principle, the green ``simulation'' part in Fig.~\ref{willsch_fig_scheme} can be completely replaced by a real digital QC once available and working.
    \item \textbf{Classical Post-Processing} (yellow): Find the largest denominator $r<N$ such that $j/2^t\approx k/r$ using a continued fraction expansion\footnote{The continued fraction expansion is a systematic method that yields successive approximations $k_0/r_0, k_1/r_1, \ldots$ with increasing denominators $r_0<r_1<\cdots$ to an arbitrary real number (cf.~e.g.~Ref.~\citen{ekert1996quantumalgorithms}).}. 
    \item \textbf{Factor Extraction} (yellow): Compute $\mathrm{gcd}(a^{\lfloor r/2\rfloor}\pm 1,N)$\footnote{We note that this expression can be computed efficiently classically, because $\mathrm{gcd}(y,N) = \mathrm{gcd}(y\,\mathrm{mod}\,N,N)$ for all $y$ and the modular exponentiation $a^x\,\mathrm{mod}\,N$ can be computed with the square-and-multiply algorithm.} which will---with sufficiently high probability---yield one of the factors $p$ or $q$.
\end{enumerate}
The proof why the algorithm works is beyond the scope of this article. However, it is important to understand that there is a certain probability that Shor's algorithm fails (even on ideal digital QCs, in part also due to the probabilistic nature of the QC model itself, as is the case for many other standard quantum algorithms~\cite{NielsenChuang}; see Appendix A.2 of Ref.~\citen{Willsch2023Shor} for more information). This motivates us to perform large-scale simulations of Shor's factoring algorithm on JUWELS Booster, to obtain a practical estimate of the success probability---i.e., what \emph{sufficiently high probability} in point 4 above means (see also Refs.~\citen{Fowler2004ShorWithImpreciseRotationGates,Nam2012ShorPerformanceBandedQFT,Nam2013ShorPerformanceBandedQFT,Nam2013ShorSimulationShort,Nam2014ShorSimulationSummary,Nam2018ShorSymmetryBoost} for related endeavours).

\subsection{Large-Scale Simulations}
\label{willsch_sec_simulations}

\begin{figure}[t]
\begin{center}
\includegraphics[width=\textwidth]{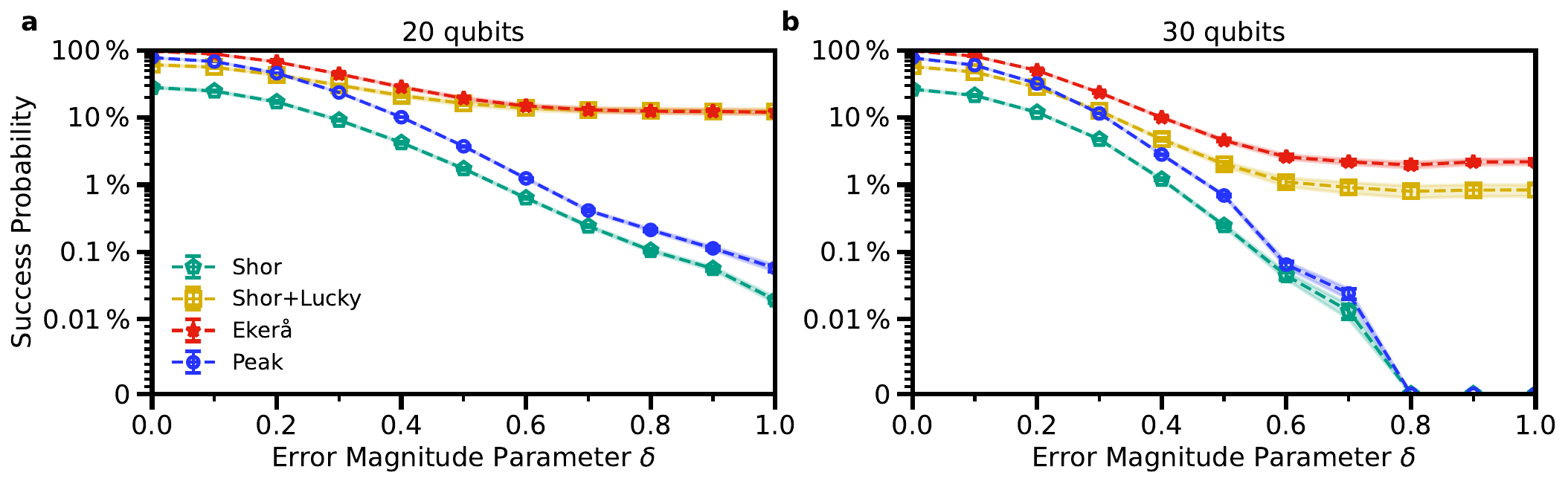}
\caption{\label{willsch_fig_errors}
Success probability of factoring on digital QCs as a function of the error magnitude $\delta$ for \textbf{a}, 20 qubits factoring 19-bit semiprimes, and \textbf{b}, 30 qubits factoring 29-bit semiprimes.
Markers denote the mean success probability over \textbf{a} 500 and \textbf{b} 1000 simulated factoring problems for each $\delta$ in four different cases: 
(i) \texttt{Shor} (green pentagons) corresponds to Shor's original factoring procedure~\cite{shor1994factoring,shor1997algorithm},
(ii) \texttt{Shor+Lucky} (yellow squares) includes the unexpected lucky cases, in which the factorisation works in practise even though the theoretical requirements~\cite{Willsch2023Shor} are not met,
(iii) \texttt{Eker{\aa}} (red stars) denotes the success probability when using the best-known classically efficient post-proccessing procedures~\cite{Ekera2022OnTheSuccessProbabilityOfQuantumOrderFindingShor,Ekera2021OnCompletelyFactoringAnyIntegerShor} on the measured bitstrings,
(iv) \texttt{Peak} (blue circles) indicates only the probability to observe a peak in Shor's bitstring-output distribution~\cite{Willsch2023Shor} that is the actual theoretical quantity studied in Cai's proof~\cite{cai2023shorsAlgorithmFailsInThePresenceOfNoise}.
At $\delta=0$, the success probabilities are between $25$--$100\,\%$, in agreement with Ref.~\citen{Willsch2023Shor} for the no-error scenarios.
Note the change from linear to logarithmic scale at $0.01\,\%$ on the vertical axes.
Shaded areas and error bars indicate the unbiased standard error of the mean.
Lines are guides to the eye.
}
\end{center}
\end{figure}

Theoretical estimates of the success probability for Shor's factoring algorithm (as described in Sec.~\ref{willsch_sec_shor}) are usually very pessimistic and amount to only a few percent~\cite{Willsch2023Shor}.
We have designed a digital QC simulation~\cite{shorgpu} of Shor's algorithm (see Fig.~\ref{willsch_fig_scheme}, green part) to evaluate the practical performance for over $60\,000$ factoring problems, with a surprising result: There are many so-called \textbf{lucky cases} in which the factorisation is successful, even though Shor's algorithm is, according to theory, not expected to work. Furthermore, we have posed the challenge of factoring, on a real QPU, a non-trivial semiprime larger than the number $N=549\,755\,813\,701=712\,321\times771\,781$ that we have factored by executing Shor's algorithm on a simulated QC.

We remark that the wall-clock time that this simulation takes actually grows only linearly with the number of qubits, due to the high degree of parallelism. However, the space complexity is exponential, as simulating the ${L+1}$-qubit quantum computer requires at least $16\times2^{L+1}$ bytes of memory~\cite{DeRaedt2007MassivelyParallel,DeRaedt2018MassivelyParallel}. Specifically, \texttt{shorgpu}~\cite{shorgpu}, as well as universal QC simulators like JUQCS--G~\cite{Willsch2021JUQCSGQAOA}, require doubling the number of GPUs with every additional qubit.

A nice benefit of a large-scale QC simulation is that it allows the study of \textbf{classical and quantum errors}, which affect any QPU device with various orders of magnitude. Of particular interest is an error model proposed by Cai in Ref.~\citen{cai2023shorsAlgorithmFailsInThePresenceOfNoise}, for which one can formally prove that Shor's factoring algorithm fails.
This error model is expressed in terms of an error magnitude parameter $\delta$\footnote{The error parameter $\delta$ used here and in Ref.~\citen{shorgpu} corresponds to the \emph{global magnitude parameter} $\epsilon$ in Ref.~\citen{cai2023shorsAlgorithmFailsInThePresenceOfNoise}, which expresses Gaussian noise on each rotation gate $R$ in the quantum circuit of Shor's algorithm (see Refs.~\citen{Willsch2023Shor}, \citen{shorgpu}, and~\citen{cai2023shorsAlgorithmFailsInThePresenceOfNoise} for more information). Specifically, the faulty rotation gate is defined as $\widetilde{R_k}=\mathrm{diag}(1, e^{2\pi i(1+\delta r)/2^k})$ where $r$ is a normally distributed random number.}. Cai's proof can be seen as formal support for the common viewpoint that for large-scale factoring on a digital QC to work, quantum error correction~\cite{NielsenChuang} would be required.

Figure~\ref{willsch_fig_errors} shows results for the success probability as a function of the error magnitude $\delta$. We see that from 20 to 30 qubits (panels \textbf{a} and \textbf{b}, respectively), the success probability for Shor's algorithm (green diamonds) indeed drops towards zero for errors with $\delta\ge0.8$. Interestingly, however, when including the lucky cases (yellow squares) or efficient classical post-processing\footnote{The parameters for Eker{\aa}'s post-processing procedures~\cite{Ekera2022OnTheSuccessProbabilityOfQuantumOrderFindingShor,Ekera2021OnCompletelyFactoringAnyIntegerShor} are the same as in Ref.~\citen{Willsch2023Shor}, i.e., $(B,c,k,\varsigma)=(L,1,100,1)$. They are such that the success probability without error ($\delta=0$) is increased to above $95\,\%$ to observe the dependence on $\delta$. Note that for the small numbers being factorised, it would be possible to increase the post-processing parameters further to achieve success probabilities close to one even in the presence of arbitrarily large errors.} (red stars), the success probability converges to a non-negligible, finite value. Even though this finite value might decrease exponentially when increasing the number of qubits---in agreement with Cai's proof---it is thus conceivable that the challenge of \emph{limited quantum speedup}~\cite{Ronnow2014DefiningQuantumSpeedup} posed in Ref.~\citen{Willsch2023Shor} may be met without the above-mentioned requirement for quantum error correction.

\subsection{Future Perspectives}
\label{willsch_sec_alternative}

The quantum circuit in Fig.~\ref{willsch_fig_scheme} needs $L+1$ qubits to factor an $L$-bit semiprime. However, on a digital QPU, the individual quantum gates usually need to be compiled into realisable one- and two-qubit gates. This is expected to yield quantum circuits with $2L$ to $2L+3$ qubits~\cite{Beauregard2003ShorWith2nplus3Qubits,TakahashiShor2nplus2,Haner2017ShorWith2nplus2QubitsToffoli,Gidney2018factoringShorWith2nplus1,kahanamokumeyer2024fastquantumintegermultiplication} (or $1.5L$ qubits with a trick~\cite{Zalka2006ShorWithFewerQubits}). As these qubits need to perform almost perfectly, a quantum error correction overhead can raise the required number of physical qubits dramatically. For instance, for the factorisation of 2048-bit RSA integers, \emph{several millions of physical qubits} are currently anticipated~\cite{Gidney2021HowToFactor2048RSAShor}.

Hence, over the past decades, there have been many algorithmic developments and alternative ideas to solve the IFP on digital QCs, often preserving the theoretical concept of an exponential speedup over current algorithms. 
In particular, the Eker{\aa}-H{\aa}stad scheme~\cite{Ekera2017FactoringWithDiscreteLogarithm} makes use of another algorithm invented by Shor, namely the discrete logarithm quantum algorithm~\cite{shor1994factoring,shor1997algorithm}. The advantage of this scheme is that it yields a roughly $75\,\%$ shorter quantum circuit\footnote{This means that $t\approx1.5L$ in Fig.~\ref{willsch_fig_scheme} would suffice instead of $t\approx2L$. In this context, it is also worth mentioning Regev's multidimensional variants of Shor's algorithm~\cite{regev2024efficientQuantumFactoringAlgorithm,ragavan2024ImprovedRegevQuantumFactoringAlgorithm,ekera2023extendingRegevsFactoringAlgorithmToDiscreteLogarithms}, which also yield an asymptotically shorter quantum circuit.}. These optimisations, however, do not directly reduce the number of qubits. 

Fascinatingly, Chevignard, Fouque, and Schrottenloher managed~\cite{Chevignard2024ReducingQubitsInShorsAlgorithmFrom2ntonover2} to combine the Eker{\aa}-H{\aa}stad scheme with a hash function technique~\cite{MaySchlieper2022CompressingShor} to obtain quantum circuits using between $0.5L$ and \emph{less than $L$} qubits to factorise $L$-bit RSA integers (see Table 3 in Ref.~\citen{Chevignard2024ReducingQubitsInShorsAlgorithmFrom2ntonover2}). It is exciting to see what further research along these lines can bring.\clearpage

\section{Analog Quantum Computers}
\label{willsch_sec_analogqc}

Like a digital QC, an analog QC is a machine consisting of individual qubits. However, in contrast to digital QCs, the qubits in an analog QC are not individually and arbitrarily controllable. Instead, after programming an analog QC, the qubits typically undergo a natural evolution for a certain time, and at the end each qubit is measured, yielding a classical-bit state. 
Note that this does not mean that analog QCs cannot be universal---in fact, one can prove a polynomial equivalence\cite{Aharonov2008AdiabaticQuantumComputationIsEquivalentToUniversalQC,AlbashLidar2018AdiabaticQuantumComputation,Imoto2024UniversalQuantumComputationWithDWave} to universal digital QCs\footnote{However, this polynomial equivalence cannot be implemented on most currently existing analog QCs due to technical limitations. For instance, the analog QC would need to support 3-local terms or 6-dimensional quantum digits~\cite{Aharonov2008AdiabaticQuantumComputationIsEquivalentToUniversalQC}, other so-called \emph{non-stoquastic} properties (cf.~Ref.~\citen{AlbashLidar2018AdiabaticQuantumComputation} for a comprehensive review), or successive back-and-forth annealing~\cite{Imoto2024UniversalQuantumComputationWithDWave}.}.

Analog QCs are easier to manufacture than digital QCs, mostly due to the more relaxed requirements on individual qubit control. Therefore, much larger analog QPU systems have been built to date. D-Wave has manufactured superconducting quantum annealers with over 5600 qubits, one of which is located in Europe---the JUPSI system hosted at JSC---and a QPU with over 7000 qubits is in development~\cite{dwave2022Advantage2}. The companies Pasqal and QuEra build analog QCs based on neutral atoms, with qubit numbers ranging from 196~\cite{Scholl2021Pasqal196qubits} to 256~\cite{wurtz2023quera256neutralatomqubits} up to 828~\cite{pichard2024rearrangementsingleatoms2000site}, and the California Institute of Technology reports 6100 coherent atomic qubits~\cite{manetsch2024tweezerarray6100qubits}.

\subsection{Factorisation on Quantum Annealers}
\label{willsch_sec_methods}

Quantum annealers are designed to solve optimisation problems. In particular, the D-Wave Advantage QPU addresses the Quadratic Unconstrained Binary Optimisation (QUBO) problem, defined as the minimisation $\min\limits_{x_i=0,1} E(x_0,x_1,\ldots,x_{n-1})$ of the cost function
\begin{equation}
  E(x_0,x_1,\ldots,x_{n-1}) = \sum_{i=0}^{n-1} a_i x_i 
  + \sum_{i < j}^{n-1} b_{ij} x_i x_j\,.
  \label{willsch_eq_qubo}
\end{equation}
Here, $n$ is the number of qubits, $x_i=0,1$ are the binary problem variables that are represented by qubits on the QPU, and $a_i$ and $b_{ij}$ are the real-valued programmable biases and couplers of the qubits, respectively.

\begin{figure}[t]
\begin{center}
\includegraphics[width=\textwidth]{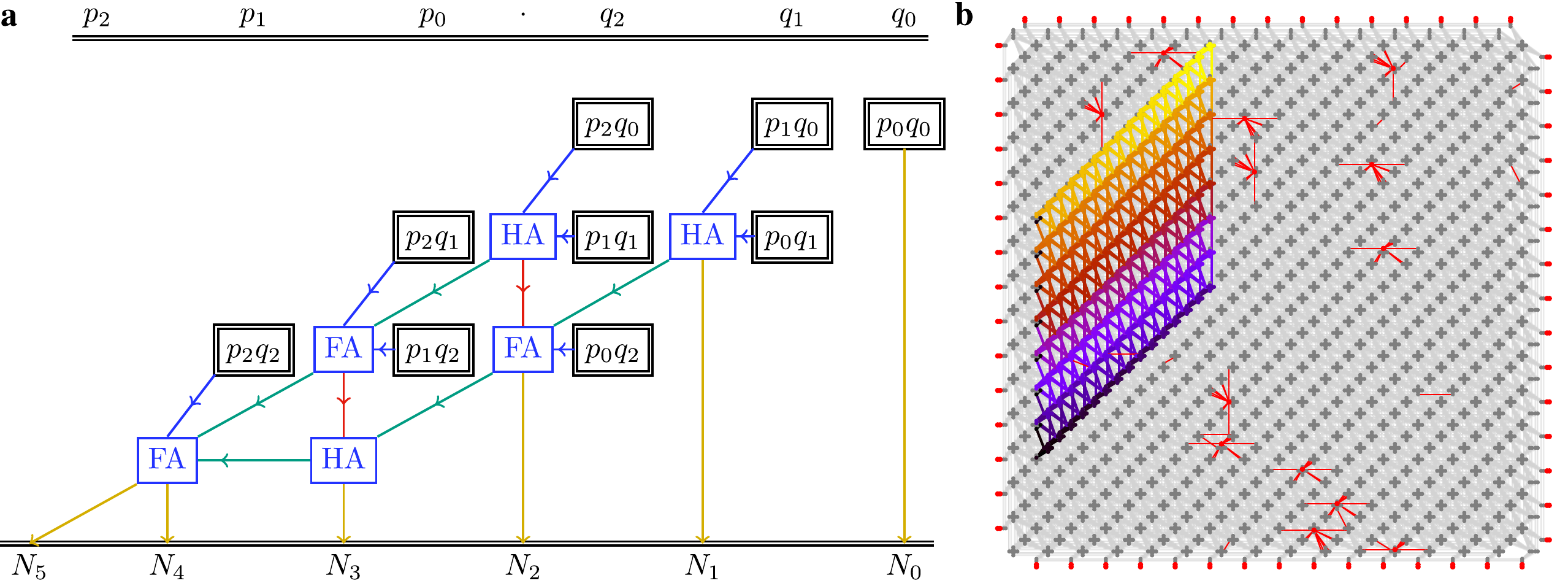}
\caption{\label{willsch_fig_methods}
Visualisation of two factorisation methods on a D-Wave quantum annealer. 
\textbf{a}, $3\times 3$-bit multiplication table for the MC method. The bits of $p$ and $q$ are connected with Boolean AND, HA, and FA logic gates. Arrows indicate immediate ancilla qubits representing \texttt{and} (blue), \texttt{sum} (red), and \texttt{carry} (green) bits.
\textbf{b}, Embedding of a $15\times8$-bit multiplier onto the D-Wave Advantage 4.1 QPU. Nodes (edges) represent the 5627 physical qubits (40277 couplers) on the QPU. Red lines indicate qubits and couplers that exist in the full underlying Pegasus graph but not on the QPU (by construction or due to fabrication defects). 
When this multiplier is used for the factorisation of e.g.~$N=3\,548\,021$; the bits of the factor $101010100000111_2$ ($10100011_2$) correspond to the vertically (diagonally) connected unit cells.
}
\end{center}
\end{figure}

To solve the IFP on quantum annealers, we therefore have to represent the solution to the IFP as the minimum of Eq.~(\ref{willsch_eq_qubo}). The most common approach is to use the qubits $x_0,x_1,\ldots\in\{0,1\}$ to represent the unknown bits of the factors $p$ and $q$. We express $p$ ($q$) using $l_p$ ($l_q$) bits\footnote{Since $l_p$ and $l_q$ have to be fixed, one would usually start with $l_p\approx l_q\approx L/2$ (where $L$ is the bit length of the semiprime $N$ to factor) and then start decreasing $l_p$ and increasing $l_q$ until a factor is found. Note that this only incurs a polynomial overhead.}. Since $N$ is odd---otherwise finding a factor would be trivial---we know that the least significant bits of $p$ and $q$ are 1. Furthermore, since $l_p$ and $l_q$ are fixed, we can set the most significant bits to one. The binary encoding of $p$ and $q$ thus reads
\begin{eqnarray}
  \label{willsch_eq_p}
  p &=& 1p_{l_p^*}p_{l_p^*-1}\cdots p_{2}p_11\,, \\
  \label{willsch_eq_q}
  q &=& 1q_{l_q^*}p_{l_q^*-1}\cdots q_{2}q_11\,,
\end{eqnarray}
where $l_p^*=l_p-2$ and $l_q^*=l_q-2$ count the number of unknown bits $l=l_p^*+l_q^*$. Given this encoding, we consider three methods to obtain a QUBO cost function $E(x_0,x_1,\ldots,x_{n-1})$ with $n$ qubits $(x_0,x_1,\ldots,x_{n-1}) = (p_1,\ldots,p_{l_p^*},q_1,\ldots,q_{l_q^*},\ldots)$:
\begin{enumerate}
    \item \textbf{Direct Method}~\cite{Peng2008QuantumAdiabaticAlgorithmForFactorization,Schaller2010FactoringAdiabaticQuantumAloorithm,Jiang2018QuantumAnnealingForPrimeFactorization}: An obvious cost function to minimise is $f(p,q) = (N-pq)^2$, as its minimum $f(p,q)=0$ is attained if and only if $N=p\times q$. However, when inserting the binary representations in Eqs.~(\ref{willsch_eq_p}) and (\ref{willsch_eq_q}) into this cost function, one obtains higher-than-quadratic terms between the qubits. To solve this problem, one uses a reduction technique that yields $n_\mathrm{reduction}$ additional so-called ancilla qubits to obtain a cost function of the form of Eq.~(\ref{willsch_eq_qubo})\footnote{In Ref.~\citen{Hanussek2024BachelorThesis}, the direct method is equal to the \emph{Modified Multiplication Table (MMT) method}~\cite{Jiang2018QuantumAnnealingForPrimeFactorization} in the limit of maximum block size, where there are no carry variables. For smaller block sizes, the ancilla qubits would consist of both $n_\mathrm{reduction}$ qubits from the quadratic reductions and $n_\mathrm{carry}$ qubits for carry bits in the multiplication table. In this article, we only consider the direct method as its performance was found to be superior to MMT with smaller block sizes~\cite{Hanussek2024BachelorThesis}.}. We thus need $n=l+n_\mathrm{reduction}$ qubits.
    \item \textbf{Multiplication Circuit Method}~\cite{Andriyash2016BoostingIntegerFactorizationDWave} (MC Method): A complimentary approach is to write out the binary product $1p_{l_p^*}p_{l_p^*-1}\cdots p_{2}p_11\times1q_{l_q^*}p_{l_q^*-1}\cdots q_{2}q_11$ in a \emph{long multiplication table}. Between all unknown bits, one can then identify Boolean AND, half-adder (HA), and full-adder (FA) gates (see Fig.~\ref{willsch_fig_methods}a). For each such gate, one can find a QUBO cost function that attains its minimum if and only if the Boolean logic gate is satisfied. The sum of all these cost functions then yields the final cost function Eq.~(\ref{willsch_eq_qubo}). We remark that also this method incurs additional ancilla qubits representing intermediate ``and'', ``sum'' and ``carry'' bits such that $n=l+n_\mathrm{and}+n_\mathrm{sum}+n_\mathrm{carry}$.
    \item \textbf{Controlled Full-Adder Method}~\cite{ding2023FactoringOnDWaveLocallyStructuredEmbedding} (CFA Method): Both direct and MC methods need many couplers $b_{ij}$ between the qubits in Eq.~(\ref{willsch_eq_qubo}). However, on the D-Wave Advantage QPU, one qubit is only coupled to 15 other qubits on average. When more connections between qubits are required than physically exist on the QPU, one has to perform a heuristic \emph{embedding}~\cite{Choi2008Embedding} step, by which multiple physical qubits are connected to represent a single logical problem variable. Such an embedding step is often found to hamper the performance of analog QCs~\cite{Willsch2021BenchmarkAdvantage}. The CFA method is a clever extension of the MC method, in which each Boolean logic gate can be directly embedded onto the qubits of the QPU (see Fig.~\ref{willsch_fig_methods}b). Finding such \emph{custom embeddings} is very often the key to successfully solve larger problems on analog QCs.
\end{enumerate}
Further details about each method are given in Ref.~\citen{Hanussek2024BachelorThesis} and supporting data and open-source code can be found in Ref.~\citen{jupsifactoring}.

A complimentary approach to solving the IFP is based on Schnorr's algorithm~\cite{Schnorr1990,Schnorr2021}. Here one does not encode the factors $p$ and $q$ directly in terms of qubits (cf.~Eqs.~(\ref{willsch_eq_p}) and (\ref{willsch_eq_q})). Instead, the IFP is first mapped to a lattice problem, which is then mapped to a QUBO problem. Interestingly, this QUBO problem may need very few qubits, and the approach has been used to factorise up to 100-bit numbers~\cite{Yan2022FactoringIntegersWithSublinearResources,Wang2024Chinese50bitFactoringRecord,hegade2023digitizedcounterdiabaticquantumfactorization,Tesoro2024QuantumInspiredFactorization100bit,Hong2025Schnorr80BitDWave}. However, concerns about the claimed scalability of the approach have been raised~\cite{Ducas2021SchnorrGateRepository,Grebnev2023PitfallsFactoringintegersSublinearResources,Khattar2023CommentFactoringintegersSublinearResources,aboumrad2023Schnorr}, and it is an open question whether Schnorr's algorithm can really be used to address large-scale IFPs.

\subsection{Results}
\label{willsch_sec_results}

\begin{figure}[t]
\begin{center}
\includegraphics[width=\textwidth]{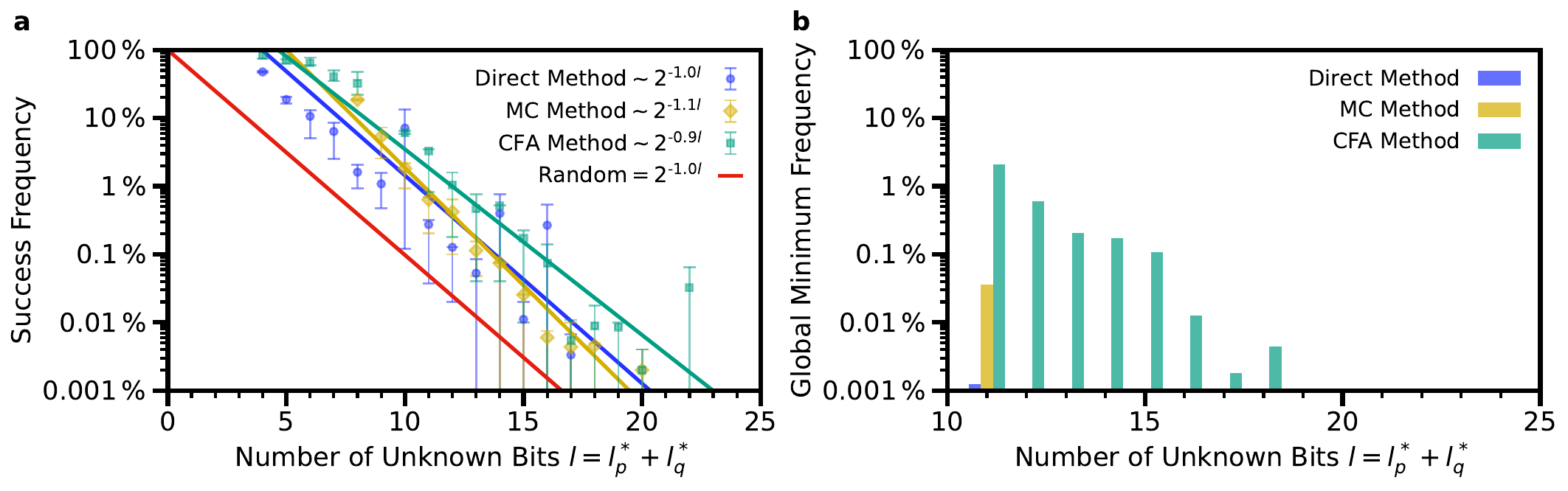}
\caption{\label{willsch_fig_dwave}
Performance of three factoring methods on analog QCs. 
\textbf{a}, Observed success frequencies as a function of the problem size, given by the number of unknown bits $l=l_p^*+l_q^*$ in the two factors $p$ and $q$. Markers represent the median success frequencies and error bars denote the $25\,\%$ and $75\,\%$ percentile for the three methods (see legend). The corresponding lines represent exponential fits $\sim2^{bl}$ to the data, with the resulting scaling exponents $b$ given in the legend. The red line represents the probability $2^{-l}$ of randomly guessing the unknown bits.
\textbf{b}, Frequency of the global minimum of the QUBO cost function, i.e., the sample in which not only the qubits representing $p$ and $q$ but also all additional ancilla qubits are correct (note that by construction of the QUBO, the solution bits representing $p$ and $q$ can be correct even though immediate carry bits are wrong).
All results have been obtained on the D-Wave Advantage QPUs 5.4 (direct method, MC method) and 4.1 (CFA method) with $\sim10000$ samples for each $N$ and about 10 randomly selected semiprimes $N$ for each $l$.
}
\end{center}
\end{figure}

We have evaluated each of the three factorisation methods for 337 randomly generated factoring problems with up to $l=22$ unknown bits. The largest factored semiprimes $N$ and the corresponding success frequencies $f$ for the three methods were:
\begin{enumerate}
    \item \textbf{Direct Method}: $N=1\,042\,441$ with $f=3.72\,\%\,,$
    \item \textbf{MC Method}: $N=1\,042\,441$ with $f=0.01\,\%\,,$
    \item \textbf{CFA Method}: $N=3\,844\,417$ with $f=0.01\,\%\,.$
\end{enumerate}
The results are shown in Fig.~\ref{willsch_fig_dwave}a. All methods show larger average success frequencies than random guessing, but the results still suggest an exponential scaling as a function of $l$.

Interestingly, the MC method with a fitted scaling exponent of $-1.1$ seems to perform asymptotically worse than random guessing\footnote{This is possible also for ideal, theoretical quantum annealing, if the annealing time is so short or the energy gap is so small that the annealing process systematically produces excited states with the wrong factors.}. We conjecture that this result is due to the requirement for additional physical qubits e.g.~from the embedding procedure.

In contrast, the custom-embedded CFA method shows a performance that seems asymptotically better than random guessing\footnote{We note that even though the fitted CFA scaling given in Fig.~\ref{willsch_fig_dwave}a is better than random guessing, further results on potentially larger QPUs might be necessary to make a statistically robust statement.}. Also, the global minimum of the cost function was found for much larger problems (see Fig.~\ref{willsch_fig_dwave}b). Although the scaling still seems exponential, a sufficiently small exponent might actually allow analog QCs to first succeed in the near-term factoring challenge posed in Ref.~\citen{Willsch2023Shor}. It is certainly interesting to see how the future Advantage2 QPU that is expected to have over 7000 qubits with most qubits coupled to 20 others~\cite{dwave2022Advantage2}---which is larger than 15 on the current JUPSI QPU---will cope with the difficult problem of factoring integers.

\section{Conclusions and Outlook}
\label{willsch_sec_conclusions}

In this article, we have studied the problem of factoring integers---one of the key problems that has fuelled the interest in quantum computing---on both digital and analog QCs. For digital QCs, we have analysed an error model for which Shor's factoring algorithm~\cite{shor1994factoring,shor1997algorithm} can be proven to fail~\cite{cai2023shorsAlgorithmFailsInThePresenceOfNoise}, and we have found that unexpected ``lucky'' factorisations~\cite{Willsch2023Shor} and sophisticated post-processing procedures~\cite{Ekera2021OnCompletelyFactoringAnyIntegerShor,Ekera2022OnTheSuccessProbabilityOfQuantumOrderFindingShor} can mitigate this effect. 

For analog QCs, we have performed experiments on a quantum annealer. Among three studied factorisation methods, we found evidence that the custom-embedded CFA method~\cite{ding2023FactoringOnDWaveLocallyStructuredEmbedding} performs absolutely and asymptotically better than random guessing, although the data still suggests an exponential scaling as a function of problem size.

Although our results suggest that either error correction on digital QCs or a new method on analog QCs would be necessary, we believe that the factorisation challenge posed in Ref.~\citen{Willsch2023Shor} might be solvable in the near term, and it will be very interesting to see whether it can first be met on a digital or an analog QC. It is conceivable---should the IFP ever be practically solvable with polynomial resources for large integers---that maybe also a triple-hybrid use~\cite{Jattana2024TripleHybrid} along with conventional supercomputers may be successful.

\section*{Acknowledgements}

The authors thank Jaka Vodeb, Paul Warburton, Jin-Yi Cai, and Martin Eker{\aa} for comments and discussions.
D.W., M.W., F.J., and K.M. gratefully acknowledge the Gauss Centre for Supercomputing e.V. (www.gauss-centre.eu) for funding this project by providing computing time on the GCS Supercomputer JUWELS~\cite{JuwelsClusterBooster} at Jülich Supercomputing Centre (JSC).
The authors gratefully acknowledge the Jülich Supercomputing Centre (https://www.fz-juelich.de/ias/jsc) for funding this project by providing computing time on the D-Wave Advantage™ System JUPSI through the Jülich UNified Infrastructure for Quantum computing (JUNIQ).
D.W. and M.W. acknowledge support from the project JUNIQ that has received funding from the German Federal Ministry of Education and Research (BMBF) and the Ministry of Culture and Science of the State of North Rhine-Westphalia.

\bibliographystyle{nic}
\bibliography{willsch_database}

\begin{thebibliography}{10}

\bibitem{Bressoud}
D.~M. Bressoud,
 {\em {Factorization and Primality Testing}},
 Springer, New York, USA, 1989.

\bibitem{Lehman1974FactoringLargeIntegers}
R.~S. Lehman,
 {\em {Factoring large integers}},
 Math. Comput., {\bf 28}, 637, 1974.

\bibitem{Boudot2020FactoringRSA250}
F.~Boudot, P.~Gaudry, A.~Guillevic, N.~Heninger, E.~Thom{\'e}, and
  P.~Zimmermann,
 {\em {Comparing the Difficulty of Factorization and Discrete Logarithm: A
  240-Digit Experiment}},
  in: {Advances in Cryptology -- CRYPTO 2020}, Daniele Micciancio and Thomas
  Ristenpart, (Eds.), p.~62, Springer International Publishing, Cham, 2020.

\bibitem{lenstra1993developmentOfTheNumberFieldSieve}
A.~K. Lenstra and H.~W. Lenstra,
 {\em {The development of the number field sieve}},
 Lecture Notes in Mathematics. Springer, Berlin, Heidelberg, 1993.

\bibitem{Kleinjung2010FactorizeRSA768Modulus}
T.~Kleinjung, K.~Aoki, J.~Franke, A.~K. Lenstra, E.~Thom{\'e}, J.~W. Bos,
  P.~Gaudry, A.~Kruppa, P.~L. Montgomery, D.~A. Osvik, H.~te~Riele,
  A.~Timofeev, and P.~Zimmermann,
 {\em {Factorization of a 768-Bit RSA Modulus}},
  in: {Advances in Cryptology -- CRYPTO 2010}, Tal Rabin, (Ed.), p. 333,
  Springer, Berlin, Heidelberg, 2010.

\bibitem{RSA1978RSA}
R.~L. Rivest, A.~Shamir, and L.~Adleman,
 {\em {A Method for Obtaining Digital Signatures and Public-key
  Cryptosystems}},
 Commun. ACM, {\bf 21}, 120, 1978.

\bibitem{shor1994factoring}
P.~W. Shor,
 {\em {Algorithms for quantum computation: discrete logarithms and factoring}},
  in: {Proceedings 35th Annual Symposium on Foundations of Computer Science},
  p. 124, Santa Fe, NM, USA, 1994.

\bibitem{ekert1996quantumalgorithms}
A.~Ekert and R.~Jozsa,
 {\em {Quantum computation and Shor's factoring algorithm}},
 Rev. Mod. Phys., {\bf 68}, 733, 1996.

\bibitem{shor1997algorithm}
P.~W. Shor,
 {\em {Polynomial-Time Algorithms for Prime Factorization and Discrete
  Logarithms on a Quantum Computer}},
 SIAM J. Comput., {\bf 26}, 1484, 1997.

\bibitem{Vandersypen2001ExperimentalShorNMR}
L.~M.~K. Vandersypen, M.~Steffen, G.~Breyta, C.~S. Yannoni, M.~H. Sherwood, and
  I.~L. Chuang,
 {\em {Experimental realization of Shor{\textquotesingle}s quantum factoring
  algorithm using nuclear magnetic resonance}},
 Nature, {\bf 414}, 883, 2001.

\bibitem{MartinLopez2012ShorExperimentQubitRecycling}
E.~Mart{\'{\i}}n-L{\'{o}}pez, A.~Laing, T.~Lawson, R.~Alvarez, X.-Q. Zhou, and
  J.~L. O{\textquotesingle}Brien,
 {\em {Experimental realization of Shor{\textquotesingle}s quantum factoring
  algorithm using qubit recycling}},
 Nat. Photonics, {\bf 6}, 773, 2012.

\bibitem{Monz2016ShorOnTrappedIons}
T.~Monz, D.~Nigg, E.~A. Martinez, M.~F. Brandl, P.~Schindler, R.~Rines, S.~X.
  Wang, I.~L. Chuang, and R.~Blatt,
 {\em {Realization of a scalable Shor algorithm}},
 Science, {\bf 351}, 1068, 2016.

\bibitem{Amico2019ShorOnIBMQ}
M.~Amico, Z.~H. Saleem, and M.~Kumph,
 {\em {Experimental study of Shor's factoring algorithm using the IBM Q
  Experience}},
 Phys. Rev. A, {\bf 100}, 012305, 2019.

\bibitem{Smolin2013OversimplifyingShorFactoring}
J.~A. Smolin, G.~Smith, and A.~Vargo,
 {\em {Oversimplifying quantum factoring}},
 Nature, {\bf 499}, 163, 2013.

\bibitem{Willsch2023Shor}
D.~Willsch, M.~Willsch, F.~Jin, H.~De~Raedt, and K.~Michielsen,
 {\em {Large-Scale Simulation of Shor{'}s Quantum Factoring Algorithm}},
 Mathematics, {\bf 11}, 4222, 2023.

\bibitem{scholten2024IBMAssessingBenefitsAndRisksOfQuantumComputing}
T.~L. Scholten, C.~J. Williams, D.~Moody, M.~Mosca, W.~Hurley, W.~J. Zeng,
  M.~Troyer, and J.~M. Gambetta,
 {\em {Assessing the Benefits and Risks of Quantum Computers}},
 arXiv:2401.16317, 2024.

\bibitem{Peng2008QuantumAdiabaticAlgorithmForFactorization}
X.~Peng, Z.~Liao, N.~Xu, G.~Qin, X.~Zhou, D.~Suter, and J.~Du,
 {\em {Quantum Adiabatic Algorithm for Factorization and Its Experimental
  Implementation}},
 Phys. Rev. Lett., {\bf 101}, 220405, 2008.

\bibitem{Schaller2010FactoringAdiabaticQuantumAloorithm}
G.~Schaller and R.~Sch\"{u}tzhold,
 {\em {The Role of Symmetries in Adiabatic Quantum Algorithms}},
 Quantum Inf. Comput., {\bf 10}, 109, 2010.

\bibitem{Andriyash2016BoostingIntegerFactorizationDWave}
E.~Andriyash, Z.~Bian, F.~Chudak, M.~Drew-Brook, A.~D. King, W.~G. Macready,
  and A.~Roy,
 {\em {Boosting integer factoring performance via quantum annealing offsets}},
 Tech. {R}ep., D-Wave Systems Inc, Burnaby, BC, Canada, 2016,
 {14-1002A-B}.

\bibitem{Dridi2017PrimeFactorizationDWave}
R.~Dridi and H.~Alghassi,
 {\em {Prime factorization using quantum annealing and computational algebraic
  geometry}},
 Sci. Rep., {\bf 7}, 43048, 2017.

\bibitem{Jiang2018QuantumAnnealingForPrimeFactorization}
S.~Jiang, K.~A. Britt, A.~J. McCaskey, T.~S. Humble, and S.~Kais,
 {\em {Quantum Annealing for Prime Factorization}},
 Sci. Rep., {\bf 8}, 17667, 2018.

\bibitem{Peng2019FactoringLargeIntegersDWave}
W.~Peng, B.~Wang, F.~Hu, Y.~Wang, X.~Fang, X.~Chen, and C.~Wang,
 {\em {Factoring larger integers with fewer qubits via quantum annealing with
  optimized parameters}},
 Sci. China Phys. Mech. Astron., {\bf 62}, 60311, 2019.

\bibitem{Mengoni2020BreakingRSAWithDWave2000Q}
R.~Mengoni, D.~Ottaviani, and P.~Iorio,
 {\em {Breaking RSA Security With A Low Noise D-Wave 2000Q Quantum Annealer:
  Computational Times, Limitations And Prospects}},
 arXiv:2005.02268, 2020.

\bibitem{Wang2020PrimeFactorizationParemeterOptimizationIsingAnnealer}
B.~Wang, F.~Hu, H.~Yao, and C.~Wang,
 {\em {Prime factorization algorithm based on parameter optimization of Ising
  model}},
 Sci. Rep., {\bf 10}, 7106, 2020.

\bibitem{Lanthaler2023ReversibleParityGatesForIntegerFactorization}
M.~Lanthaler, B.~E. Niehoff, and W.~Lechner,
 {\em {Scalable set of reversible parity gates for integer factorization}},
 Commun. Phys., {\bf 6}, 1--8, 2023.

\bibitem{ding2023FactoringOnDWaveLocallyStructuredEmbedding}
J.~Ding, G.~Spallitta, and R.~Sebastiani,
 {\em {Effective prime factorization via quantum annealing by modular
  locally-structured embedding}},
 Sci. Rep., {\bf 14}, 1, 2024.

\bibitem{NielsenChuang}
M.~A. Nielsen and I.~L. Chuang,
 {\em {Quantum Computation and Quantum Information: 10th Anniversary Edition}},
 Cambridge University Press, New York, 2010.

\bibitem{Castelvecchi2023IBM1121qubits}
D.~Castelvecchi,
 {\em {IBM releases first-ever 1,000-qubit quantum chip}},
 Nature, {\bf 624}, 238, 2023.

\bibitem{google2024quantumerrorcorrection105qubits}
{Google Quantum AI},
 {\em {Quantum error correction below the surface code threshold}},
 arXiv:2408.13687, 2024.

\bibitem{Bluvstein2024NeutralAtomLogicalQuantumProcessor}
D.~Bluvstein, S.~J. Evered, A.~A. Geim, S.~H. Li, H.~Zhou, T.~Manovitz,
  S.~Ebadi, M.~Cain, M.~Kalinowski, and D.~Hangleiter~et al,
 {\em {Logical quantum processor based on reconfigurable atom arrays}},
 Nature, {\bf 626}, 58, 2024.

\bibitem{daSilva2024QuantinuumH2}
M.~P. da~Silva, C.~Ryan-Anderson, J.~M. Bello-Rivas, A.~Chernoguzov, J.~M.
  Dreiling, C.~Foltz, F.~Frachon, J.~P. Gaebler, T.~M. Gatterman, and
  L.~Grans-Samuelsson et~al.,
 {\em {Demonstration of logical qubits and repeated error correction with
  better-than-physical error rates}},
 arXiv:2404.02280, 2024.

\bibitem{decross2024QuantinuumH2Upgrade}
M.~DeCross, R.~Haghshenas, M.~Liu, E.~Rinaldi, J.~Gray, Y.~Alexeev, C.~H.
  Baldwin, J.~P. Bartolotta, M.~Bohn, and E.~Chertkov et~al.,
 {\em {The computational power of random quantum circuits in arbitrary
  geometries}},
 arXiv:2406.02501, 2024.

\bibitem{Ronkko2024IQMSpark}
J.~R{\ifmmode\ddot{o}\else\"{o}\fi}nkk{\ifmmode\ddot{o}\else\"{o}\fi},
  O.~Ahonen, V.~Bergholm, A.~Calzona, A.~Geresdi, H.~Heimonen, J.~Heinsoo,
  V.~Milchakov, S.~Pogorzalek, M.~Sarsby, M.~Savytskyi, S.~Seegerer,
  F.~{\ifmmode\check{S}\else\v{S}\fi}imkovic, P.~V. Sriluckshmy, P.~T. Vesanen,
  and M.~Nakahara,
 {\em {On-premises superconducting quantum computer for education and
  research}},
 EPJ Quantum Technol., {\bf 11}, 1, 2024.

\bibitem{Piltz2014eleQtronQPU}
C.~Piltz, T.~Sriarunothai, A.~F. Var{\ifmmode\acute{o}\else\'{o}\fi}n, and
  C.~Wunderlich,
 {\em {A trapped-ion-based quantum byte with 10{-}5 next-neighbour
  cross-talk}},
 Nat. Commun., {\bf 5}, 1, 2014.

\bibitem{shorgpu}
D.~Willsch,
 {\em {\texttt{shorgpu}: Simulation of Shor's algorithm with the semiclassical
  Fourier transform using multiple GPUs and MPI}},
 \url{https://jugit.fz-juelich.de/qip/shorgpu.git}, 2023.

\bibitem{Fowler2004ShorWithImpreciseRotationGates}
A.~G. Fowler and L.~C.~L. Hollenberg,
 {\em {Scalability of Shor's algorithm with a limited set of rotation gates}},
 Phys. Rev. A, {\bf 70}, 032329, 2004.

\bibitem{Nam2012ShorPerformanceBandedQFT}
Y.~S. Nam and R.~Bl\"umel,
 {\em {Performance scaling of Shor's algorithm with a banded quantum Fourier
  transform}},
 Phys. Rev. A, {\bf 86}, 044303, 2012.

\bibitem{Nam2013ShorPerformanceBandedQFT}
Y.~S. Nam and R.~Bl\"umel,
 {\em {Scaling laws for Shor's algorithm with a banded quantum Fourier
  transform}},
 Phys. Rev. A, {\bf 87}, 032333, 2013.

\bibitem{Nam2013ShorSimulationShort}
Y.~S. Nam and R.~Bl\"umel,
 {\em {Streamlining Shor's algorithm for potential hardware savings}},
 Phys. Rev. A, {\bf 87}, 060304, 2013.

\bibitem{Nam2014ShorSimulationSummary}
Y.~S. Nam and R.~Bl\"umel,
 {\em {Robustness and performance scaling of a quantum computer with respect to
  a class of static defects}},
 Phys. Rev. A, {\bf 88}, 062310, 2013.

\bibitem{Nam2018ShorSymmetryBoost}
Y.~S. Nam and R.~Bl\"umel,
 {\em {Symmetry boost of the fidelity of Shor factoring}},
 Phys. Rev. A, {\bf 97}, 052311, 2018.

\bibitem{Ekera2022OnTheSuccessProbabilityOfQuantumOrderFindingShor}
M.~Eker\r{a},
 {\em {On the Success Probability of Quantum Order Finding}},
 ACM Trans. Quantum Comput., {\bf 5}, 11, 2024.

\bibitem{Ekera2021OnCompletelyFactoringAnyIntegerShor}
M.~Eker{\aa},
 {\em {On completely factoring any integer efficiently in a single run of an
  order-finding algorithm}},
 Quantum Inf. Process., {\bf 20}, 205, 2021.

\bibitem{cai2023shorsAlgorithmFailsInThePresenceOfNoise}
J.-Y. Cai,
 {\em {Shor{'}s algorithm does not factor large integers in the presence of
  noise}},
 Sci. China Inf. Sci., {\bf 67}, 1, 2024.

\bibitem{DeRaedt2007MassivelyParallel}
K.~{De Raedt}, K.~Michielsen, H.~{De Raedt}, B.~Trieu, G.~Arnold, M.~Richter,
  {\relax Th}.~Lippert, H.~Watanabe, and N.~Ito,
 {\em {Massively parallel quantum computer simulator}},
 Comput. Phys. Commun., {\bf 176}, 121, 2007.

\bibitem{DeRaedt2018MassivelyParallel}
H.~{De Raedt}, F.~Jin, D.~Willsch, M.~Willsch, N.~Yoshioka, N.~Ito, S.~Yuan,
  and K.~Michielsen,
 {\em {Massively parallel quantum computer simulator, eleven years later}},
 Comput. Phys. Commun., {\bf 237}, 47, 2019.

\bibitem{Willsch2021JUQCSGQAOA}
D.~Willsch, M.~Willsch, F.~Jin, K.~Michielsen, and H.~{De Raedt},
 {\em {GPU}-accelerated simulations of quantum annealing and the quantum
  approximate optimization algorithm},
 Comput. Phys. Commun., {\bf 278}, 108411, 2022.

\bibitem{Ronnow2014DefiningQuantumSpeedup}
T.~F. R{\o}nnow, Z.~Wang, J.~Job, S.~Boixo, S.~V. Isakov, D.~Wecker, J.~M.
  Martinis, D.~A. Lidar, and M.~Troyer,
 {\em {Defining and detecting quantum speedup}},
 Science, {\bf 345}, 420, 2014.

\bibitem{Beauregard2003ShorWith2nplus3Qubits}
S.~Beauregard,
 {\em {Circuit for Shor's algorithm using 2n+3 qubits}},
 Quantum Inf. Comput., {\bf 3}, 175, 2003.

\bibitem{TakahashiShor2nplus2}
Y.~Takahashi and N.~Kunihiro,
 {\em {A quantum circuit for Shor's factoring algorithm using 2n+2 qubits}},
 Quantum Inf. Comput., {\bf 6}, 0184, 2006.

\bibitem{Haner2017ShorWith2nplus2QubitsToffoli}
T.~H\"aner, M.~Roetteler, and K.~M. Svore,
 {\em {Factoring using 2n+2 qubits with Toffoli based modular multiplication}},
 Quantum Inf. Comput., {\bf 17}, 0673, 2017.

\bibitem{Gidney2018factoringShorWith2nplus1}
C.~Gidney,
 {\em {Factoring with n+2 clean qubits and n-1 dirty qubits}},
 arXiv:1706.07884, 2018.

\bibitem{kahanamokumeyer2024fastquantumintegermultiplication}
G.~D. Kahanamoku-Meyer and N.~Y. Yao,
 {\em {Fast quantum integer multiplication with zero ancillas}},
 arXiv:2403.18006, 2024.

\bibitem{Zalka2006ShorWithFewerQubits}
C.~Zalka,
 {\em {Shor's algorithm with fewer (pure) qubits}},
 arXiv:quant-ph/0601097, 2006.

\bibitem{Gidney2021HowToFactor2048RSAShor}
C.~Gidney and M.~Eker{\aa{}},
 {\em {How to factor 2048 bit {RSA} integers in 8 hours using 20 million noisy
  qubits}},
 {Quantum}, {\bf 5}, 433, 2021.

\bibitem{Ekera2017FactoringWithDiscreteLogarithm}
M.~Eker{\aa} and J.~H{\aa}stad,
 {\em {Quantum Algorithms for Computing Short Discrete Logarithms and Factoring
  RSA Integers}},
  in: {Post-Quantum Cryptography}, Tanja Lange and Tsuyoshi Takagi, (Eds.), p.
  347, Springer International Publishing, Cham, 2017.

\bibitem{regev2024efficientQuantumFactoringAlgorithm}
O.~Regev,
 {\em {An Efficient Quantum Factoring Algorithm}},
 arXiv:2308.06572, 2024.

\bibitem{ragavan2024ImprovedRegevQuantumFactoringAlgorithm}
S.~Ragavan and V.~Vaikuntanathan,
 {\em {Space-Efficient and Noise-Robust Quantum Factoring}},
  in: {Advances in Cryptology -- CRYPTO 2024}, Leonid Reyzin and Douglas
  Stebila, (Eds.), p. 107, Springer Nature Switzerland, Cham, 2024.

\bibitem{ekera2023extendingRegevsFactoringAlgorithmToDiscreteLogarithms}
M.~Eker{\aa} and J.~G{\"a}rtner,
 {\em {Extending Regev's Factoring Algorithm to Compute Discrete Logarithms}},
  in: {Post-Quantum Cryptography}, Markku-Juhani Saarinen and Daniel
  Smith-Tone, (Eds.), p. 211, Springer Nature Switzerland, Cham, 2024.

\bibitem{Chevignard2024ReducingQubitsInShorsAlgorithmFrom2ntonover2}
C.~Chevignard, P.-A. Fouque, and A.~Schrottenloher,
 {\em {Reducing the Number of Qubits in Quantum Factoring}},
 {Cryptology ePrint Archive, Paper 2024/222}, 2024.

\bibitem{MaySchlieper2022CompressingShor}
A.~May and L.~Schlieper,
 {\em {Quantum Period Finding is Compression Robust}},
 IACR Trans. Symmetric Cryptol., {\bf 2022}, 183, 2022.

\bibitem{Aharonov2008AdiabaticQuantumComputationIsEquivalentToUniversalQC}
D.~Aharonov, W.~van Dam, J.~Kempe, Z.~Landau, S.~Lloyd, and O.~Regev,
 {\em {Adiabatic Quantum Computation is Equivalent to Standard Quantum
  Computation}},
 {SIAM} J. Comput., {\bf 37}, 166, 2007.

\bibitem{AlbashLidar2018AdiabaticQuantumComputation}
T.~Albash and D.~A. Lidar,
 {\em {Adiabatic quantum computation}},
 Rev. Mod. Phys., {\bf 90}, 015002, 2018.

\bibitem{Imoto2024UniversalQuantumComputationWithDWave}
T.~Imoto, Y.~Susa, R.~Miyazaki, T.~Kadowaki, and Y.~Matsuzaki,
 {\em {Universal quantum computation using quantum annealing with the
  transverse-field Ising Hamiltonian}},
 arXiv:2402.19114, 2024.

\bibitem{dwave2022Advantage2}
C.~McGeoch, Pau Farr{\'{e}}, and K.~Boothby,
 {\em {The D-Wave Advantage2 Prototype}},
 Tech. {R}ep., D-Wave Systems Inc, Burnaby, BC, Canada, 2022,
 {14-1063A-A}.

\bibitem{Scholl2021Pasqal196qubits}
P.~Scholl, M.~Schuler, H.~J. Williams, A.~A. Eberharter, D.~Barredo, K.-N.
  Schymik, V.~Lienhard, L.-P. Henry, T.~C. Lang, T.~Lahaye, A.~M.
  L{\ifmmode\ddot{a}\else\"{a}\fi}uchli, and A.~Browaeys,
 {\em {Quantum simulation of 2D antiferromagnets with hundreds of Rydberg
  atoms}},
 Nature, {\bf 595}, 233, 2021.

\bibitem{wurtz2023quera256neutralatomqubits}
J.~Wurtz, A.~Bylinskii, B.~Braverman, J.~Amato-Grill, S.~H. Cantu, F.~Huber,
  A.~Lukin, F.~Liu, P.~Weinberg, J.~Long, S.-T. Wang, N.~Gemelke, and
  A.~Keesling,
 {\em {Aquila: QuEra's 256-qubit neutral-atom quantum computer}},
 arXiv:2306.11727, 2023.

\bibitem{pichard2024rearrangementsingleatoms2000site}
G.~Pichard, D.~Lim, E.~Bloch, J.~Vaneecloo, L.~Bourachot, G.-J. Both,
  G.~M{\'{e}}riaux, S.~Dutartre, R.~Hostein, J.~Paris, B.~Ximenez, A.~Signoles,
  A.~Browaeys, T.~Lahaye, and D.~Dreon,
 {\em {Rearrangement of single atoms in a 2000-site optical tweezers array at
  cryogenic temperatures}},
 arXiv:2405.19503, 2024.

\bibitem{manetsch2024tweezerarray6100qubits}
H.~J. Manetsch, G.~Nomura, E.~Bataille, K.~H. Leung, X.~Lv, and M.~Endres,
 {\em {A tweezer array with 6100 highly coherent atomic qubits}},
 arXiv:2403.12021, 2024.

\bibitem{Hanussek2024BachelorThesis}
P.~Hanussek,
 {\em {Comparison of Factoring Algorithms on the D-Wave Quantum Annealer}},
 \url{https://doi.org/10.34734/FZJ-2024-05254}, 2024.

\bibitem{Choi2008Embedding}
V.~Choi,
 {\em {Minor-embedding in adiabatic quantum computation: I. The parameter
  setting problem}},
 Quantum Inf. Process., {\bf 7}, 193, 2008.

\bibitem{Willsch2021BenchmarkAdvantage}
D.~Willsch, M.~Willsch, C.~D. Gonzalez~Calaza, F.~Jin, H.~{De Raedt},
  M.~Svensson, and K.~Michielsen,
 {\em {B}enchmarking {A}dvantage and {D-Wave 2000Q} quantum annealers with
  exact cover problems},
 Quantum Inf. Process., {\bf 21}, 141, 2022.

\bibitem{jupsifactoring}
P.~Hanussek,
 {\em {JupsiFactoring: Factoring on the D-Wave Quantum Annealer}},
 \url{https://jugit.fz-juelich.de/qip/jupsifactoring}, 2024.

\bibitem{Schnorr1990}
C.{-}P. Schnorr,
 {\em Factoring Integers and Computing Discrete Logarithms via Diophantine
  Approximation},
  in: Advances In Computational Complexity Theory, Proceedings of a {DIMACS}
  Workshop, New Jersey, USA, December 3-7, 1990, Jin{-}Yi Cai, (Ed.), vol.~13
  of {\em {DIMACS} Series in Discrete Mathematics and Theoretical Computer
  Science}, pp. 171--181, {DIMACS/AMS}. 1990.

\bibitem{Schnorr2021}
C.{-}P. Schnorr,
 {\em {Fast Factoring Integers by SVP Algorithms, corrected}},
 Cryptology ePrint Archive, Paper 2021/933, 2021.

\bibitem{Yan2022FactoringIntegersWithSublinearResources}
B.~Yan, Z.~Tan, S.~Wei, H.~Jiang, W.~Wang, H.~Wang, L.~Luo, Q.~Duan, Y.~Liu,
  W.~Shi, Y.~Fei, X.~Meng, Y.~Han, Z.~Shan, J.~Chen, X.~Zhu, C.~Zhang, F.~Jin,
  H.~Li, C.~Song, Z.~Wang, Z.~Ma, H.~Wang, and G.-L. Long,
 {\em {Factoring integers with sublinear resources on a superconducting quantum
  processor}},
 arXiv:2212.12372, 2022.

\bibitem{Wang2024Chinese50bitFactoringRecord}
C.~Wang, Q.-D. Wang, C.-L. Hong, Q.-Y. Hu, and Z.~Pei,
 {\em Research on Public-Key Cryptosystem Attack Algorithm Based on D-Wave
  Quantum Annealing},
 Chin. J. Comput., {\bf 47}, 13015, 2024.

\bibitem{hegade2023digitizedcounterdiabaticquantumfactorization}
N.~N. Hegade and E.~Solano,
 {\em Digitized-counterdiabatic quantum factorization},
 arXiv:2301.11005, 2023.

\bibitem{Tesoro2024QuantumInspiredFactorization100bit}
M.~Tesoro, I.~Siloi, D.~Jaschke, G.~Magnifico, and S.~Montangero,
 {\em Quantum inspired factorization up to 100-bit RSA number in polynomial
  time},
 arXiv:2410.16355, 2024.

\bibitem{Hong2025Schnorr80BitDWave}
C.~Hong, Z.~Pei, Q.~Wang, S.~Yang, J.~Yu, and C.~Wang,
 {\em {Quantum attack on RSA by D-Wave Advantage: a first break of 80-bit
  RSA}},
 Sci. China Inf. Sci., {\bf 68}, 1, 2025.

\bibitem{Ducas2021SchnorrGateRepository}
L.~Ducas,
 ``Schnorrgate: A framework for exploring lattice-based cryptography'',
 \url{https://github.com/lducas/SchnorrGate}, 2021.

\bibitem{Grebnev2023PitfallsFactoringintegersSublinearResources}
S.~V. Grebnev, M.~A. Gavreev, E.~O. Kiktenko, A.~P. Guglya, A.~R. Efimov, and
  A.~K. Fedorov,
 {\em Pitfalls of the Sublinear QAOA-Based Factorization Algorithm},
 IEEE Access, {\bf 11}, 134760, 2023.

\bibitem{Khattar2023CommentFactoringintegersSublinearResources}
T.~Khattar and N.~Yosri,
 {\em A comment on "Factoring integers with sublinear resources on a
  superconducting quantum processor"},
 arXiv:2307.09651, 2023.

\bibitem{aboumrad2023Schnorr}
W.~Aboumrad, D.~Widdows, and A.~Kaushik,
 {\em Quantum and Classical Combinatorial Optimizations Applied to
  Lattice-Based Factorization},
 arXiv:2308.07804, 2023.

\bibitem{Jattana2024TripleHybrid}
M.~S. Jattana,
 {\em {Quantum annealer accelerates the variational quantum eigensolver in a
  triple-hybrid algorithm}},
 Phys. Scr., {\bf 99}, 095117, 2024.

\bibitem{JuwelsClusterBooster}
{J}\"ulich Supercomputing~Centre,
 {\em {JUWELS Cluster and Booster: Exascale Pathfinder with Modular
  Supercomputing Architecture at Juelich Supercomputing Centre}},
 J. of Large-Scale Res. Facil., {\bf 7}, A183, 2021.

\end{thebibliography}

\end{document}